\documentclass[a4paper,11pt]{article}
\usepackage{latexsym}
\usepackage{amsfonts} 
\usepackage[mathscr]{euscript}
\usepackage{amsmath,mathrsfs,bm,amssymb,color}

\title{\textsf{Upper bounds on  the charge  susceptibility of 
many-electron systems coupled to the quantized radiation field }}
\date{\empty}

\author{
Tadahiro Miyao\\ 
 {\it Department of Mathematics,}
{\it Hokkaido University,}\\
{\it Sapporo 060-0810, Japan}\\
E-mail:
 miyao@math.sci.hokudai.ac.jp
}

\newcommand{\one}{{\mathchoice {\rm 1\mskip-4mu l} {\rm 1\mskip-4mu l}
{\rm 1\mskip-4.5mu l} {\rm 1\mskip-5mu l}}}
\newcommand{\h}{\mathfrak{h}}

\newcommand{\ex}{\mathrm{e}}
\newcommand{\D}{\mathrm{dom}}

\newcommand{\im}{\mathrm{i}}
\newcommand{\Fock}{\mathfrak{F}}

\newcommand{\dG}{d\Gamma}

\newcommand{\AFock}{\mathfrak{F}_{\mathrm{a}}}
\newcommand{\la}{\langle}
\newcommand{\ra}{\rangle}
\newcommand{\Tr}{\mathrm{Tr}}

\newcommand{\BbbR}{\mathbb{R}}
\newcommand{\BbbN}{\mathbb{N}}
\newcommand{\BbbZ}{\mathbb{Z}}

\newcommand{\BbbC}{\mathbb{C}}
\newcommand{\vepsilon}{\varepsilon}

\newcommand{\Hf}{H_{\mathrm{f}}}

\newcommand{\no}{\nonumber \\}

\newcommand{\mb}{\mathbf}

\newcommand{\bfx}{x}
\newcommand{\bfy}{y}
\newcommand{\bfk}{k}
\newcommand{\bfA}{A}
\newcommand{\bfr}{r}

\begin{document}

\newtheorem{define}{Definition}[section]
\newtheorem{Thm}[define]{Theorem}
\newtheorem{Prop}[define]{Proposition}
\newtheorem{lemm}[define]{Lemma}
\newtheorem{rem}[define]{Remark}
\newtheorem{assum}{Condition}
\newtheorem{example}{Example}
\newtheorem{coro}[define]{Corollary}

\maketitle

\begin{abstract}
We extend the Kubo-Kishi theorem concerning the charge susceptibility  of the Hubbard model
 in the following way: 
(i) The electron-photon interaction   is taken into account.
(ii) Not only on-site  but also general 
 Coulomb repulsions are  considered.  
\end{abstract} 

\section{Introduction and the results}

In order to study the origin of ferromagnetism,
 Gutzwiller \cite{Gutzwiller},  Hubbard \cite{Hubbard}  and   Kanamori \cite{Kanamori}
 proposed a tight binding model of electrons with on-site 
Coulomb interaction, which is called the Hubbard model nowadays.
If there is no Coulomb interaction,
the model demonstrates paramagnetism due to the Pauli exclusion principle.
Therefore, in 1960's, researchers had a common question 
whether  it is  possible to transform  the paramagnetism to
ferromagnetism by taking the Coulomb
 interaction into account. The  first rigorous result about ferromagnetism in the Hubbard model was given by
Nagaoka \cite{Nagaoka}.  He showed  that ferromagnetism appears in the ground state when there  is 
 exactly a single hole and the strength of the Coulomb interaction is infinite.
Since then, the Hubbard model has been widely recognized as a simplified
model for rigorous study of ferromagnetism.
For a review of the rigorous results, we refer to  \cite{Tasaki}.

In 1989, Lieb showed that the Hubbard model on a connected bipartite lattice at half filling  exhibits 
ferrimagnetism by applying the spin reflection positivity \cite{Lieb}.
In 1990, Kubo and Kishi \cite{KuboKishi} showed the absence of  charge long range order in the Hubbard system
 on the bipartite lattice by using the method of Dyson, Lieb and Simon \cite{DLS}.
Their result can be regarded as a finite temperature version of Lieb's theorem.

In this paper, we rigorously study a crystal coupled to the quantized radiation field.
Our Hamiltonian is the Hubbard model with an electron-photon interaction
term. Besides,  not only on-site but also general  Coulomb repulsions 
are  considered.   
The purpose of this paper is to extend the Kubo-Kishi theorem to such a
generalized case.
We note  that the influence of the classical electromagnetic field was
 studied by Lieb \cite{Lieb2}.  In contrast we take account
of effects of the quantized electromagnetic field. 
Also we remark that a  similar model was considered  in \cite{GMP, GMP2}.   The effects of
electron-photon interaction in graphen were investigated  by an exact renormalization group analysis. 
In this paper, we will not  restrict ourself to the honeycomb lattice. 
Our method can be extend to  general bipartite  lattices straightforwardly.

\subsection{Background: The Hubbard model } 
First of all, we review some  known results for the Hubbard model.
For each $d\in \{1,2,3\}$ and $\ell\in \BbbN$ even, let 
\begin{align}
 \Lambda= [-\ell/2, \ell/2)^d\cap \mathbb{Z}^d.
\end{align} 
In what follows, we regard $\Lambda$ as a simple cubic torus.
The Hubbard model is given by
\begin{align} 
H_{\mathrm{H}}=&\sum_{ {x, y\in \Lambda, |x-y|=1}\atop{ \sigma=\uparrow, \downarrow}}
\big(-t\big)
c_{\bfx\sigma}^*
 c_{\bfy\sigma}
+ \frac{1}{2}\sum_{x \in \Lambda}
 U_0(n_{x}-\one)( n_{x}-\one).
\end{align}
$H_{\mathrm{H}}$ lives  in the Hilbert space 
\begin{align}
\mathfrak{E} =\Fock_{\mathrm{a}}(\ell^2(\Lambda)\oplus \ell^2(\Lambda)), 
\end{align} 
where $\Fock_{\mathrm{a}}(\h)$ is the fermionic Fock space over $\h$
defined by $\AFock(\h)=\oplus_{n=0}^{\infty}\wedge^n \h$.  
$\wedge^n \h$ is the $n$-fold anti-symmetric tensor product of $\h$. 
$c_{x\sigma}$ and $c_{x\sigma}^*$ are the fermionic annihilation- and
creation operators respectively. The following anti-commutation
relations hold:
\begin{align}
&\{ c_{x\sigma}, c_{x'\sigma'}^*\}=\delta_{\sigma \sigma'} \delta_{x x'},\\
&\{c_{x\sigma}, c_{x'\sigma'}\}=0=\{c_{x\sigma}^*, c_{x'\sigma'}^*\}.
\end{align} 
$n_{x}=n_{x\uparrow}+n_{x\downarrow}$ is the  number operator at
vertex $x$ with $n_{x\sigma}=c_{x\sigma}^* c_{x\sigma}$.
$U_0$ is the strength of the on-site Coulomb interaction. $t>0$ is the
nearest-neighbor hopping integral.

Let $\delta n_x:=n_x-\one$. Set
\begin{align}
\widetilde{\delta n}_p=|\Lambda|^{-1/2} \sum_{x\in \Lambda}\ex^{-\im
 x\cdot p} \delta n_x.
\end{align} 
The charge susceptibility is defined by  
\begin{align}
\chi_{\beta, \mathrm{H}}(p)=\lim_{\ell \to \infty}\beta\big(\widetilde{\delta
 n}_{-p},\widetilde{\delta n}_p\big)_{\beta, \Lambda, \mathrm{H}},
\end{align} 
where $( \cdot, \cdot )_{\beta, \Lambda, \mathrm{H}}$ is the Duhamel
two-point function given  by 
\begin{align}
(A, B)_{\beta, \Lambda, \mathrm{H}}&=Z_{\beta, \Lambda, \mathrm{H}}^{-1} \int_0^1 ds \Tr\Big[
\ex^{-s \beta H_{\mathrm{H}}}A \, \ex^{-(1-s)\beta H_{\mathrm{H}}} B
\Big],\\
Z_{\beta, \Lambda, \mathrm{H}}&=\Tr\big[
\ex^{-\beta H_{\mathrm{H}}}
\big].
\end{align} 
We employ the thermal average with respect to the
grand canonical Gibbs state at inverse temperature $\beta$.
For any $\beta$ and $\Lambda$, we can check that the thermal average
 density of the system satisfies $\la n_o\ra_{\beta, \Lambda,
 \mathrm{H}}:=Z_{\beta, \Lambda, \mathrm{H}}^{-1} \Tr[n_o \ex^{-\beta H_{\mathrm{H}}}]=1$, that is,
 the system at half-filling is considered.

In \cite{KuboKishi}, Kubo and Kishi proved the following:

\begin{Thm}\label{KKThm}
Assume  $U_0>0$. Then  one has 
\begin{align}
\chi_{\beta, \mathrm{H}}(p)\le U_0^{-1}
\end{align} 
 for all $p\in [-\pi, \pi]^d$. 
\end{Thm} 

Using the Falk-Bruch inequality and Theorem \ref{KKThm}, we conclude the absence of the charge
long-range order.

\subsection{Many-electron system coupled to the 
  radiation field}

For  $L\in \BbbN$ even, set 
\begin{align}
V=[-L/2, L/2)^3,\hspace{1cm} V^*=\Big(\frac{2\pi}{L}\BbbZ\Big)^3.
\end{align} 
In this paper, we always assume that $L\ge \ell$.  Thus we can regard
$\Lambda$ as a subset of $V$ by the following manner:
\begin{align}
\Lambda \equiv 
\begin{cases}
\{(x, 0, 0)\in V\, |\, x\in \Lambda\} & d=1\\
\{(x, 0)\in V\, |\, x\in \Lambda\} & d=2 \\
\Lambda & d=3. 
\end{cases}
\end{align} 
We are interested in the interacting electron-photon system. The total
Hamiltonian is 
\begin{align}
H=&\sum_{ {x, y\in \Lambda,\ |x-y|=1,}\atop{ \sigma=\uparrow, \downarrow}}
\big(-t\big)
\exp\Bigg\{
\im e \int_{C_{\bfx\bfy}} d\bfr\cdot  \bfA(\bfr)\Bigg\}c_{\bfx\sigma}^*
 c_{\bfy\sigma}\\
&+ \frac{1}{2} \sum_{x, y \in \Lambda}
 U(x-y)(n_{x}-\one )(n_{y}-\one )
+ \sum_{\lambda=1,2}\sum_{k\in V^*}\omega(\bfk)a(\bfk, \lambda)^* a(\bfk, \lambda)\\
=:& H_{\mathrm{e}-\mathrm{p}}+H_{\mathrm{e}-\mathrm{e}}+\Hf. \label{DefHamiltonian}
\end{align}
$H$ acts in the Hilbert space 
\begin{align}
\mathfrak{E} \otimes \mathfrak{P},
\end{align} 
where  $\mathfrak{P}=\Fock_{\mathrm{s}}(L^2(V^*\times \{1,2\}))$. 
$\Fock_{\mathrm{s}}(\h)$ is the bosonic Fock space defined by 
$\Fock_{\mathrm{s}}(\h)=\oplus_{n=0}^{\infty}\otimes_{\mathrm{s}} \h$ ,
 where $\otimes_{\mathrm{s}}^n \h$ is the $n$-fold symmetric tensor
 product of $\h$.
$a(k, \lambda)$ and $a(k, \lambda)^*$ are the bosonic annihilation- and
 creation operators respectively. These satisfy the canonical
 commutation relations: 
\begin{align}
&[a(k, \lambda), a(k',
 \lambda')^*]=\delta_{\lambda \lambda'}\delta_{kk'},\\
&[a(k, \lambda), a(k',
 \lambda')]=0=[a(k, \lambda)^*, a(k',
 \lambda')^*].
\end{align} 

The quantized vector potential is given by 
\begin{align}
  \bfA(\bfx)&=|V|^{-1/2}\sum_{\lambda=1,2}\sum_{k\in V^*} \frac{\chi_{\kappa}(\bfk)}{\sqrt{2\omega(\bfk)}} \vepsilon(\bfk,
 \lambda)
\Big(
a(\bfk, \lambda)\, \ex^{\im \bfk\cdot \bfx}+a(\bfk, \lambda)^* \ex^{-\im \bfk\cdot \bfx}
\Big).
\end{align} 
The form factor $\chi_{\kappa}$ is the indicator function of the ball of radius
$\kappa<\infty$. The dispertion relation is chosen to be $\omega(k)=|k|$ for
$k\in V^*\backslash \{0\}$, $\omega(0)=m_0$ with $0<m_0<2\pi/L$.\footnote{The modification at $k=0$ is needed in order to guarantee that $\ex^{-\beta
H}$ is trace class. However when we take the limit $\ell\to \infty$ at the
final stage of our arguments, our conclusions are independent  of the value
of $m_0$.}
$C_{xy}$ is a piecewise smooth curve from $x$ to $y$.
The polarization vectors are denoted by $\vepsilon(k,
\lambda)=(\vepsilon_1(k ,\lambda), \vepsilon_2(k, \lambda),
\vepsilon_3(k, \lambda)),\ \lambda=1,2$.
Together with $k/|k|$, they form a basis, which for concreteness is
taken as 
\begin{align}
\vepsilon(k, 1)=\frac{(k_2, -k_1, 0)}{\sqrt{k_1^2+k_2^2}},\ \ \
 \vepsilon(k, 2)=\frac{k}{|k|}\wedge \vepsilon(k, 1).
\end{align} 
For convenience, we put $\vepsilon(k, \lambda)=0$ if $(k_1, k_2)=(0,
0)$.
The charge of a single electron is denoted by $e$.
 $U(x)$ is a real-valued function on $\BbbZ^d  $ such that $\|U\|_{\infty}:=\max_{x\in
\BbbZ^d} |U(x)|<\infty$ and $U(-x)=U(x)$.
It is trivial that $H$ is self-adjoint and bounded from below.

We assume the following:
\begin{itemize}
\item[{\bf (A. 1)}] $U\in \ell^1(\BbbZ^d)$.
\item[{\bf (A. 2)}]  For all $\ell>0$, it holds that $\hat{U}_{\Lambda}(p)\ge 0$, where
	     $\hat{f}_{\Lambda}(p)=\sum_{x\in \Lambda}\ex^{-\im x\cdot p}f(x)$. 
\end{itemize}

As before the charge susceptibility is defined by 

\begin{align}
\chi_{\beta}(p)=\lim_{\ell \to \infty}\beta\big(\widetilde{\delta
 n}_{-p},\widetilde{\delta n}_p \big)_{\beta, \Lambda},
\end{align} 
where $(\cdot, \cdot)_{\beta, \Lambda}$ is the Duhamel two-point
function associated with $H$.

The following theorem is an extension of Theorem \ref{KKThm}.
\begin{Thm}\label{SusUpper}
Assume {\bf (A. 1)} and {\bf (A. 2)}. For each $p\in [-\pi, \pi]^d$
 such that $\hat{U}(p)>0$, one has 
\begin{align}
\chi_{\beta}(p)\le \hat{U}(p)^{-1}.
\end{align}
\end{Thm}

\begin{coro}\label{Absence}
Assume {\bf (A. 1)} and {\bf (A. 2)}. In addition, assume that  there exists a
 constant $u_0>0$ such that   $\hat{U}(p)\ge
 u_0$
for all $p\in [-\pi, \pi]^d$. Then one has 
\begin{align}
\chi_{\beta}(p)\le u_0^{-1}.
\end{align} 
Thus there is no charge long-range order.
\end{coro}

\begin{example}
{\rm
For each $U_0>0$, let $U(x)=U_0\delta_{xo}$, where $\delta_{xy}$ is the
 Kronecker delta. Then {\bf (A. 1)} and {\bf (A. 2)} are fulfilled with
 $\hat{U}_{\Lambda}(p)=U_0$. 
For all $U_0>0$,  there is no charge long-range order.
$\diamondsuit$
}
\end{example} 

\begin{example}
{\rm
For each $U_0, U_1\ge 0$, we define
\begin{align}
U(x)=\begin{cases}
 U_0 & x=o\\
U_1/2d & |x|=1\\
0 & \mbox{otherwise}
\end{cases}. 
\end{align} 
Then we see that 
$
\hat{U}_{\Lambda}(p)=(U_0-U_1)+\frac{U_1}{d}\sum_{j=1}^d(1+\cos p_j).
$
Clearly {\bf (A. 1)} and  {\bf (A. 2)} are satisfied provided $U_0\ge U_1$.
There is no charge long-range order provided that $U_0>U_1$ by Corollary \ref{Absence}. On the other
 hand,  if 
 $U_0= U_1$, then $\chi_{\beta}(p)$ could diverge at extreme points of
 $[-\pi, \pi]^d$.
$\diamondsuit$
}
\end{example}

\subsection{Organization of the paper}

The rest of this paper is organized as follows.
In Section \ref{Prel}, as a preliminary section, we construct a trace formula for the quantized 
electromagnetic field. Then, in Section \ref{Sec3}, we will prove Theorem \ref{SusUpper}.

\section{Preliminaries}\label{Prel}
\setcounter{equation}{0}

In \cite{BHL}, the Schr\"odinger representation of the radiation field
is constructed. On the other hand, Arai presented some trace formulas
associated with Gibbs states by introducing Euclidean bose field
\cite{Arai, Arai2}.
By combining the ideas in \cite{Arai, Arai2, BHL}, we will derive a
trace formula concerning the photon field in this section. 

Let $q(\cdot, \cdot)$ be a bilinear form from $\oplus^3 L^2(V)\times
\oplus^3 L^2(V)$  to $\BbbC$ defined by
\begin{align}
q(f, g)=\frac{1}{2} |V|^{-1}\sum_{k\in V^*} \big\la \hat{f}(k),
 (\one_3-|\tilde{k}\ra\la \tilde{k}|) \hat{g}(k)\big\ra_{\BbbC^3},\  f,g\in \oplus^3 L^2(V),
\end{align} 
where $\tilde{k}=k/|k|$ if $(k_1, k_2)\neq 0$, $\tilde{k}=0$ otherwise. 
 $\hat{f}(k)$ means 
$\hat{f}(k)=(\hat{f}_1(k), \hat{f}_2(k), \hat{f}_3(k))$ with 
\begin{align}
\hat{f}_j(k)=\frac{1}{\sqrt{|V|}}
\int_V dx f_j(x)\ex^{-\im x\cdot k}.
\end{align} 
 Since
$\one_3-|\tilde{k}\ra \la \tilde{k}|$ is a  positive semidefinite matrix in
$\BbbC^3$, one sees that $ q(f, f)\ge 0$. However $q$ is degenerate.
Namely   $q(f, f)= 0$ does
not imply $f=0$. Let $\mathcal{N}=\{f\in \oplus^3 L^2(V)\, |\, q(f,
f)=0\}$.  Now we define a Hilbert space by 
\begin{align}
\mathcal{H}(V)=~\overline{\oplus^3 L^2(V)/\mathcal{N}}^{\, q}.
\end{align} 
By definition, the inner product of $\mathcal{H}(V)$ satisfies
\begin{align}
\la [f], [g]\ra_{\mathcal{H}(V)}=q(f, g), \ f, g\in \oplus^3 L^2(V),
\end{align} 
where $[f] \in \oplus^3 L^2(V)/\mathcal{N}$ is the equivalence class of $f$.
Henceforth we denote $[f]$ by $f$ if no confusion occurs.

Let $U_t, t\ge 0$ be an operator defined by 
\begin{align}
\widehat{U_t f}(k) =\ex^{-t \omega(k)} \hat{f}(k),\ \ f\in
 \oplus^3 L^2(V),\  k\in V^*.
\end{align} 
Since $\|U_t f \|_{\mathcal{H}(V)} \le  \| f \|_{\mathcal{H}(V)}$
for $f\in \oplus^3 L^2(V)$,  one can extend $U_t$ to a bounded
operator on $\mathcal{H}(V)$. We also denote it by the same symbol. It
is not so hard to see that $U_t$ is a strongly continuous  one-parameter semigroup. Thus
there is a self-adjoint operator $\check{\omega}$ such that $U_t=\ex^{-t
 \check{\omega}}$. For  each $f\in \oplus^3 L^2(V) \cap \D( \check{\omega})$, we have that 
$\widehat{\check{\omega} f}(k)=\omega(k)\hat{f}(k)$. If $\gamma
>3$, then $
\sum_{k\in V^*} \omega(k)^{-\gamma}
$  is finite. Hence $\Tr_{\mathcal{H}(V)}[\check{\omega}^{-\gamma}]$ is
finite as well.

Let $\{ \phi(f)\, |\, f\in \mathcal{H}(V)\}$ be the Gaussian random
process indexed by $\mathcal{H}(V)$.  We denote by $(\mathcal{Q},
\mathcal{B}, \nu)$  the underlying probability space of the process.

For each $s\in \BbbR$,  we define an inner product $(\cdot, \cdot)_s$ on
$\D(\check{\omega}^s)$ by 
\begin{align}
(f, g)_s:= \Big\la \check{\omega}^{s/2} f, \check{\omega}^{s/2}g\Big\ra_{\mathcal{H}(V)}, \ f, g\in \D(\check{\omega}^s).
\end{align} 
For $s\ge 0$, $\mathcal{H}_s(V)=(\D(\check{\omega}^s), (\cdot,
\cdot)_s)$  becomes a Hilbert space. For $s<0$, we denote  by
$\mathcal{H}_s(V)$ the completion of $\mathcal{H}(V)$ in the norm
$\|\cdot \|_s:=(\cdot, \cdot)_s^{1/2}$.  For all $s\in \BbbR$ the dual space of
$\mathcal{H}_s(V)$ can be identified with $\mathcal{H}_{-s}(V)$ through
the bilinear form $_{-s}\la \cdot, \cdot \ra_s$ on
$\mathcal{H}_{-s}(V)\times \mathcal{H}_s(V)$ such that 
$_{-s}\la f, g\ra_s=\la f, g\ra_{\mathcal{H}(V)}$ for $f\in
\mathcal{H}(V)\cap \mathcal{H}_{-s}(V), g\in \mathcal{H}(V)\cap
\mathcal{H}_s(V)$. Fix $\gamma>3$ arbitrarily. Since $\check{\omega}^{-\gamma}$
 is in  the trace class, we can take $\mathcal{Q}=\mathcal{H}_{-\gamma}(V)$
and $\phi(f)=  {} _{-\gamma}\la \phi, f\ra_{\gamma}$ by a theorem of
 Gross \cite{Gross}.
 Let
$\mathcal{A}(f)$ be the multiplication operator by the function $
_{-\gamma}\la \phi, f\ra_{\gamma}$.
As usual we can define the Wick  product as follows:
 \begin{align}
:\mathcal{A}(f):&=\mathcal{A}(f),\\
:\mathcal{A}(f_1)\cdots
 \mathcal{A}(f_n):&=\mathcal{A}(f_1):\mathcal{A}(f_2)\cdots
 \mathcal{A}(f_n):\no
&-\sum_{j=2}^n q(f_1, f_j):\mathcal{A}(f_2)\cdots
 \widehat{\mathcal{A}(f_j)}\cdots \mathcal{A}(f_n):
\end{align} 
for $f, f_1, \dots, f_n\in \oplus^3 L^2(V)$, where $\widehat{\mathcal{A}(f_j)}$ indicates the omission of $\mathcal{A}(f_j)$.
 For each bounded operator $S$ on $\mathcal{H}_{\gamma}(V)$, its  second
quantization  $\Gamma(S)$ is defined by
\begin{align}
\Gamma(S):\mathcal{A}(f_1)\cdots
 \mathcal{A}(f_n):=:\mathcal{A}(S f_1)\cdots
 \mathcal{A}(S f_n):.
\end{align} 
Let $T$ be a  self-adjoint operator on
$\mathcal{H}_{\gamma}(V)$.
Then $\Gamma(\ex^{\im t  T})$ becomes a strongly continuous  one-parameter unitary group on
$L^2(\mathcal{Q})$. Thus there exists a unique self-adjoint operator $\dG(T)$
such that $\Gamma(\ex^{\im t T})=\ex^{\im t \dG(T)}$.

Set $\mathcal{A}_j(f):=\mathcal{A}(\oplus_{i=1}^3 \delta_{ij}f)$.
Let  $\iota$ be a unitary operator from 
$L^2(\mathcal{Q})$ onto $\mathfrak{P}$ defined by 
\begin{align}
\iota:\mathcal{A}_{j_1}(f_1)\cdots \mathcal{A}_{ j_n}(f_n):
=A_{ j_1}(f_1)_+\cdots A_{ j_n}(f_n)_+ \Omega_{\mathrm{s}},\ \ f_1,
 \dots, f_n\in L^2(\BbbR^3),
\end{align} 
where $\Omega_{\mathrm{s}}$ is the bosonic Fock vacuum and 
\begin{align}
A_{j}(f)&=A_{ j}(f)_++A_{ j}(f)_-,\\ 
\left.
\begin{array}{l}
A_{j}(f)_{+}\\
A_j(f)_-
\end{array}
\right\}
&=(2|V|)^{-1/2}
\sum_{\lambda=1,2} \sum_{k\in V^*} \vepsilon(k,
 \lambda)\hat{f}(\pm k)
\left\{
\begin{array}{l}
 a(k, \lambda)^{*}\\
a(k, \lambda)
\end{array} 
\right..
\end{align} 
Then we obtain 
\begin{align}
\iota  \overline{ A_{j}(f)}\iota^{-1}=\mathcal{A}_{ j}(f),\ \ \iota
 H_{\mathrm{f}}\iota^{-1}=d\Gamma(\check{\omega}). \label{AMulti}
\end{align} 

Let $Z_{\beta}=\Tr_{L^2(\mathcal{Q})}[\ex^{-\beta d\Gamma(\check{\omega})}]$. As is
well-known,  the Planck's formula holds:
\begin{align}
Z_{\beta}= \prod_{k\in V^*}\frac{1}{1-\ex^{-\beta \omega(k)}}.
\end{align} 

For $\beta>0$, let 
$
\mathcal{Q}_{ \beta}=C([0, \beta]; \mathcal{Q})
$ be  the space of $\mathcal{Q}$-valued continuous functions on $[0, \beta]$.
For each $\Phi\in \mathcal{Q}_{ \beta}$, we denote the value at 
$s\in [0, \beta]$ by $\Phi_s$. Then there exists a probability space
$(\mathcal{Q}_{ \beta}, \mathcal{F}_{\beta},  \nu_{\beta})$ such that $\{\Phi_s(f)\, |\,
f\in \mathcal{H}_{\gamma}(V), s\in [0, \beta]\}$  is a family of
Gaussian random variables on $(\mathcal{Q}_{\beta},
\mathcal{F}_{\beta}, \nu_{\beta})$
with covariance 
\begin{align}
\int_{\mathcal{Q}_{\beta}} \Phi_t(f)\Phi_s(g)d\nu_{\beta}(\Phi)
=\Big\la f, \big(\one-\ex^{-\beta \check{\omega}}\big)^{-1} \big(\ex^{-(\beta-|t-s|)
 \check{\omega}}+\ex^{-|t-s|\check{\omega}} \big)g\Big\ra_{\mathcal{H}(V)}.
\end{align} 
Moreover we have the following  useful formula.
\begin{lemm}\label{TraceFormula}
Let $F_1, \dots, F_n\in L^{\infty}(\mathcal{Q})$ and let
 $0\le t_1<t_2<\cdots < t_n \le \beta$. Then we have 
\begin{align}
&
\Tr_{L^2(\mathcal{Q})}
\Big[
\ex^{-t_1 d\Gamma(\check{\omega})}F_1
 \ex^{-(t_2-t_1)d\Gamma(\check{\omega})}
\cdots \ex^{-(t_n-t_{n-1}) d\Gamma(\check{\omega})}
F_n
\ex^{-(\beta-t_n) d\Gamma(\check{\omega})}
\Big]
\Bigg/ Z_{\beta}\no
=&\int_{\mathcal{Q}_{ \beta}} F_1(\Phi_{t_1})\cdots F_n(\Phi_{t_n})d\nu_{\beta}(\Phi).
\end{align} 
\end{lemm} 
{\it Proof.} See \cite{Arai, Arai2}(cf. \cite{HK}). $\Box$

\section{Proof of Theorem \ref{SusUpper}} \label{Sec3}
\setcounter{equation}{0}

Our strategy of the proof is  similar to  \cite{Miyao}.  For reader's
convenience, we provide a complete proof.

\subsection{Rewriting the Hamiltonian} \label{RewH}

\subsubsection{Expression  of the Hamiltonian in
  $(\Fock_{\mathrm{a}}\otimes \Fock_{\mathrm{a }})\otimes
  \mathfrak{P}$}

Note that
  $\mathfrak{E}=\AFock\otimes \AFock $ with $\AFock=\AFock(\ell^2(\Lambda))$. Under this
 identification, we see that 
\begin{align}
c_{x\uparrow}=c_x\otimes \one,\ \ \ c_{x\downarrow}=(-\one)^{\mathsf{N}_{\mathrm{a}}}\otimes c_x,
\end{align} 
where $c_x$ and $c_x^*$ are the fermionic annihilation- and creation
operators on $\AFock$, $\mathsf{N}_{\mathrm{a}}$ is the fermionic number
operator given by  $\mathsf{N}_{\mathrm{a}}=\sum_{x\in \Lambda}\mathsf{n}_x$ with $\mathsf{n}_x=c_x^*c_x$.
Thus our Hamiltonian becomes 
\begin{align}
H= H_{\mathrm{e-p}}+ H_{\mathrm{e-e}}\otimes \one+\one
 \otimes \one \otimes  H_{\mathrm{f}}
\end{align} 
where 
\begin{align}
H_{\mathrm{e-p}}&=-T_{+e, \uparrow} - T_{+e, \downarrow},\\
T_{\pm e, \uparrow}&= \sum_{x, y\in \Lambda,\ |x-y|=1}
t  
c_{\bfx}^*
 c_{\bfy} \otimes \one \otimes 
\exp\Bigg\{
\pm\im e \int_{C_{\bfx\bfy}} d\bfr\cdot  \bfA(\bfr)\Bigg\},\\
T_{\pm e, \downarrow}&= \sum_{x, y\in \Lambda,\ |x-y|=1}
t
\one \otimes c_{\bfx}^*
 c_{\bfy}  \otimes 
\exp\Bigg\{
\pm\im e \int_{C_{\bfx\bfy}} d\bfr\cdot  \bfA(\bfr)\Bigg\}
\end{align} 
 and 
\begin{align}
H_{\mathrm{e-e}}=\frac{1}{2}\sum_{x, y\in
 \Lambda}U(x-y)(\mathsf{n}_x\otimes\one +\one \otimes \mathsf{n}_x-\one)(
 \mathsf{n}_y\otimes \one +\one \otimes \mathsf{n}_y-\one). \label{EETensor}
\end{align} 

\subsubsection{The hole-particle transformation}

Note that  $\Lambda$
can be devided into two disjoint sets $\Lambda_e$ and $\Lambda_o$, where
$\Lambda_e=\{x\in \Lambda\, |\, x_1+x_2+\cdots +x_d\ \mbox{is even}\}$ and 
$\Lambda_o=\{x\in \Lambda\, |\, x_1+x_2+\cdots +x_d\ \mbox{is odd}\}$.
The  hole-particle transformation is a unitary operator $\mathcal{U}$ on $\mathfrak{E}$
such that  
\begin{align}
&\mathcal{U}c_{\bfx}\otimes \one \mathcal{U}^*=\gamma(\bfx)
 c_{\bfx}^*\otimes \one,\ \ \ \ 
\mathcal{U}c_{\bfx}^*\otimes \one  \mathcal{U}^*=\gamma(\bfx)
 c_{\bfx}\otimes \one,\ \ 
\mathcal{U}\one \otimes  c_{\bfx}\mathcal{U}^* =\one \otimes c_{\bfx }, \label{HoleParticle}
\end{align} 
where $\gamma(\bfx)=1$ for $\bfx\in \Lambda_e,\ \gamma(\bfx)=-1$ for
$\bfx\in \Lambda_o$.

\begin{lemm}\label{PHformula}
Let $\widehat{H}_{\mathrm{e-\mathrm{{p}}}}=
 \mathcal{U}H_{\mathrm{e}-\mathrm{p}}\mathcal{U}^*$ and
 $\widehat{H}_{\mathrm{e-e}}=\mathcal{U} H_{\mathrm{e-e}}\mathcal{U}^*$.
We have the following:
\begin{itemize}
\item[{\rm (i)}] $ \displaystyle 
\widehat{ H}_{\mathrm{e}-\mathrm{p}} =- T_{-e, \uparrow}-T_{+e, \downarrow}.
$
\item[{\rm (ii)}] 
$\displaystyle 
\widehat{H}_{\mathrm{e-e}}= 
	     \frac{1}{2} \sum_{x,y\in \Lambda}U(x-y)
	     (\mathsf{n}_x\otimes\one -\one \otimes \mathsf{n}_x)
	     (\mathsf{n}_y\otimes \one -\one \otimes \mathsf{n}_y)$.
\end{itemize} 
 \end{lemm} 
{\it Proof.} (i)
 By the definition of $\mathcal{U}$, we have 
\begin{align}
\mathcal{U} T_{+e, \uparrow} \mathcal{U}^* =
\sum_{x, y\in \Lambda,\ |x-y|=1}
t \gamma
 (\bfx)\gamma(\bfy)
c_x c_{\bfy}^*\otimes \one\otimes 
\exp\Bigg\{
\im e \int_{C_{\bfx\bfy}} d\bfr\cdot  \bfA(\bfr)\Bigg\} .
\end{align} 
Because  $\gamma(\bfx)\gamma(\bfy)=-1$ holds provided
$|x-y|=1$, we have  
\begin{align}
=&\sum_{x, y\in \Lambda,\ |x-y|=1}
t
 c_{\bfy}^*
 c_{\bfx}\otimes \one\otimes 
\exp\Bigg\{
\im e \int_{C_{\bfx\bfy}} d\bfr\cdot  \bfA(\bfr)\Bigg\}\no
=& \sum_{x, y\in \Lambda,\ |x-y|=1}
t
 c_{\bfx}^*
 c_{\bfy}\otimes \one\otimes 
\exp\Bigg\{
-\im e \int_{C_{\bfx\bfy}} d\bfr\cdot  \bfA(\bfr)\Bigg\}\no
=& T_{-e, \uparrow}.
\end{align} 
Here we used that 
$\int_{C_{\bfy\bfx}}d\bfr\cdot \bfA(\bfr)=-
\int_{C_{\bfx\bfy}}d\bfr\cdot \bfA(\bfr)$.
Similarly one obtains $\mathcal{U} 
T_{+e}\mathcal{U}^*=T_{+e}$. (ii) is obvious.
 $\Box$ 
\medskip\\

Now we arrive at  the following expression:
\begin{align}
&\mathcal{U}H \mathcal{U}^*
=\widehat{H}, \\
&\widehat{H}= \widehat{H}_{\mathrm{e-p}}+\widehat{H}_{\mathrm{e-e}}\otimes \one
 \otimes \one +\one \otimes \one \otimes H_{\mathrm{f}},
\label{TransH}
\end{align} 
where $\widehat{H}_{\mathrm{e-p}}$ and $\widehat{H}_{\mathrm{e-e}}$ are 
defined   in Lemma \ref{PHformula}.

\subsection{Gaussian domination}

For each $\mb{h}=\{h_x\}_{x\in \Lambda}\in \BbbR^{|\Lambda|}$,  let
$\widehat{H}(\mb{h})$ be the Hamiltonian $\widehat{H}$ with 
\begin{align}
H_{\mathrm{e-e}}(\mb{h})=\frac{1}{2}\sum_{x, y\in \Lambda}
 U(x-y)(\mathsf{n}_{x}\otimes \one -\one \otimes
 \mathsf{n}_{x}-h_x)(\mathsf{n}_{y}\otimes \one - \one \otimes \mathsf{n}_{y}-h_y).
\end{align}
Note that   $\widehat{H}=\widehat{H}(\mb{0})$ holds.

Under the identification $
\mathfrak{E}\otimes \mathfrak{P}=(\AFock\otimes \AFock)\otimes L^2(\mathcal{Q})
=\int_{\mathcal{Q}}^{\oplus} \AFock\otimes \AFock d\nu
$,  we have 
\begin{align}
\widehat{H}_{\mathrm{e-p}}=-\int_{\mathcal{Q}}^{\oplus} T_{-e,
 \uparrow}(\phi)d\nu(\phi)
-\int_{\mathcal{Q}}^{\oplus} T_{+e,
 \downarrow}(\phi)d\nu(\phi),
\end{align} 
where
\begin{align}
T_{\pm e, \uparrow}(\phi)&=\sum_{x, y\in \Lambda,\ |x-y|=1}t \exp\Bigg\{
\pm \im e
\int_{C_{xy}}dr\cdot \mathscr{A}(r)(\phi)
\Bigg\} c_x^* c_y\otimes \one, \label{Hopp1}\\
T_{\pm e, \downarrow}(\phi)&=\sum_{x, y\in \Lambda,\ |x-y|=1}t \exp\Bigg\{
\pm \im e
\int_{C_{xy}}dr\cdot \mathscr{A}(r)(\phi)
\Bigg\} \one \otimes c_x^* c_y,\label{Hopp2}
\end{align} 
and the vector potential $\mathscr{A}(x)$ is given by
$\mathscr{A}(x)=\mathcal{A}(\oplus^3 \rho(\cdot - x))$ with 
$ \rho=(\omega ^{-1/2} \chi_{\kappa})^{\vee}
$, see (\ref{AMulti}).  Here $\check{f}$ is the inverse Fourier transformation of $f\in
\ell^2(V^*)$.

Let $K= \widehat{H}_{\mathrm{e-p}}+d\Gamma(\check{\omega})$
and let 
\begin{align}
\mathcal{Z}_{\beta, n, \vepsilon}(\mb{h})=\Tr\Big[
\Big(
 \ex^{-\beta K/n}
\ex^{-\beta \widehat{H}_{\mathrm{e-e}}(\mb{h})/n}
\Big)^{n} \ex^{-\vepsilon d\Gamma(\check{\omega})}
\Big],\ \  n\in \BbbN,\ \  \vepsilon>0.
\end{align} 
\begin{lemm}\label{TraceLemma}
Let $T(\Phi_s) =T_{-e, \uparrow}(\Phi_s)+T_{+e, \downarrow}(\Phi_s)
$, where $T_{\pm e, \sigma}(\Phi_s)$ is given by
 (\ref{Hopp1}) and (\ref{Hopp2}) with $\mathscr{A}(r)(\Phi_s), \ \Phi\in
 \mathcal{Q}_{\beta}$. One has 
\begin{align}
&
\mathcal{Z}_{\beta, n, \vepsilon}(\mb{h})\Big/Z_{\beta+\vepsilon}
\no
=&(4\pi)^{-n|\Lambda|/2}\int_{\BbbR^{n|\Lambda|}}
\prod_{j=1}^n d\mb{k}_j \int_{\mathcal{Q}_{\beta+\vepsilon}}d\nu_{\beta+\vepsilon}(\Phi)
\ex^{-\im \sum_{j=1}^n \mb{h}\cdot \mb{k}_j}
\ex^{-\sum_{j=1}^n\mb{k}_j^2/4} \no
&\times \Tr_{\Fock_{\mathrm{a}} \otimes \Fock_{\mathrm{a}}}
\Bigg[
\prod_{j=1}^{{n}\atop{\longrightarrow}}  \Bigg\{\Bigg(
\prod_{(j-1)\beta/n}^{{j\beta/n}\atop{\longrightarrow}}\ex^{T(\Phi_s)ds}\Bigg)
\exp\Bigg\{\im \sum_{j=1}^n
\sum_{x, y\in \Lambda} \frac{\beta }{2n}k_{jx}U(x-y)(\mathsf{n}_{y}\otimes \one
 -\one \otimes \mathsf{n}_{y})
\Bigg\}
\Bigg\}
\Bigg],
\end{align} 
where 
$\displaystyle 
\prod_{j=1}^{{m}\atop{\longrightarrow}} A_j:=A_1A_2\cdots A_m,
$  the ordered product, and $\displaystyle 
\prod_{(j-1)\beta/n}^{{j\beta/n}\atop{\longrightarrow}} \ex^{T(\Phi_s) ds}
$ is the strong product integration  defined by (\ref{StrongInt}) below.
\end{lemm} 
{\it Proof.}
By the Trotter-Kato product formula, Lemma \ref{TraceFormula} and Lemma \ref{Grumm} below, we have
\begin{align}
&\mathcal{Z}_{\beta, n, \vepsilon}(\mb{h})\Big/ Z_{\beta+\vepsilon}\\
=&\frac{1}{Z_{\beta+\vepsilon}} \lim_{M_1\to \infty}\cdots \lim_{M_n \to \infty}
\Tr_{\mathfrak{H}_M}
\Bigg[
\Bigg(
\ex^{-\beta d\Gamma(\check{\omega}) /nM_1}
\ex^{-\beta \widehat{H}_{\mathrm{e-p}}/nM_1}
\Bigg)^{M_1}
\ex^{-\beta  \widehat{H}_{\mathrm{e-e}}(\mb{h})/n}
\times \no
&\hspace{2cm}\cdots\times\Bigg(
\ex^{-\beta d\Gamma (\check{\omega}) /nM_n}
\ex^{-\beta \widehat{H}_{\mathrm{e-p}}/nM_n}
\Bigg)^{M_n}
\ex^{-\beta  \widehat{H}_{\mathrm{e-e}}(\mb{h})/n}
\ex^{-\vepsilon d\Gamma(\check{\omega})}
\Bigg]\no
=&\lim_{M_1\to \infty}\cdots \lim_{M_n \to \infty}
\int_{\mathcal{Q}_{ \beta+\vepsilon}}d\nu_{\beta+\vepsilon}(\Phi)
\Tr_{\Fock_{\mathrm{a}}\otimes \Fock_{\mathrm{a}}}
\Bigg[
\Bigg(
\prod_{j=1}^{{M_1}\atop{\longrightarrow}}
\exp\Bigg\{
\frac{\beta}{nM_1} T\Big(\Phi_{ \frac{j}{n M_1} \beta}\Big)
\Bigg\}
\Bigg)
\ex^{-\beta  \widehat{H}_{\mathrm{e-e}}(\mb{h})/n}
 \no
&\hspace{2cm}
\times\Bigg(
\prod_{j=1}^{{M_2}\atop{\longrightarrow}}
\exp\Bigg\{
\frac{\beta}{nM_2} T\Big(\Phi_{ \frac{1}{n}\beta+\frac{j}{n M_1} \beta}\Big)
\Bigg\}
\Bigg)
\ex^{-\beta  \widehat{H}_{\mathrm{e-e}}(\mb{h})/n}\no
&\hspace{2cm}\cdots\times\Bigg(
\prod_{j=1}^{{M_n}\atop{\longrightarrow}}
\exp\Bigg\{\frac{\beta}{nM_n}
 T
\Big(\Phi_{
\frac{n-1}{n}\beta+\frac{j}{n M_n}\beta
}\Big
)
\Bigg\}
\Bigg)
\ex^{-\beta  \widehat{H}_{\mathrm{e-e}}(\mb{h})/n}
\Bigg].\label{TraceF}
\end{align} 
Note that $T(\Phi_s)$ is continuous in $s$ for each $\Phi\in
\mathcal{Q}_{ \beta}$. Thus the following
strong product integration  exists \cite{Dollard}:
\begin{align}
\mbox{s-}\lim_{M\to \infty}  \prod_{j=1}^{{M}\atop{\longrightarrow}}
\exp\Bigg\{
\frac{\beta}{nM} T
\Big(
\Phi_{s+\frac{j}{nM}\beta}
\Big)
\Bigg\}
=:\prod_{s}^{{s+\frac{\beta}{n}}\atop{\longrightarrow}}
 \ex^{T(\Phi_s)ds}, \label{StrongInt}
\end{align} 
where $\mbox{s-}\lim$ means the strong limit. Thus 
we see that the R.H.S. of (\ref{TraceF}) converges to
\begin{align} 
\int_{\mathcal{Q}_{\beta+\vepsilon}}d\nu_{\beta+\vepsilon}(\Phi)
\Tr_{\Fock_{\mathrm{a}}\otimes \Fock_{\mathrm{a}}}
\Bigg[
\Bigg(
\prod_{0}^{{\frac{\beta}{n}}\atop{\longrightarrow}}
\ex^{
T(\Phi_s)ds
}
\Bigg)
\ex^{-\beta  \widehat{H}_{\mathrm{e-e}}(\mb{h})/n}\times \no
\cdots\times\Bigg(
\prod_{\frac{n-1}{n}\beta}^{{\beta}\atop{\longrightarrow}}
\ex^{
T(\Phi_s)ds
}
\Bigg)
\ex^{-\beta  \widehat{H}_{\mathrm{e-e}}(\mb{h})/n}
\Bigg].
\end{align} 
Note that since $U(x-y)$ is a positive semidefinite matrix, it holds
that 
\begin{align}
&\ex^{-\beta \widehat{H}_{\mathrm{e-e}}(\mb{h})/n}\no
=&(4\pi)^{-|\Lambda|/2}\int_{\BbbR^{|\Lambda|}} d\mb{k}
\ex^{-\im \mb{h}\cdot \mb{k}}\ex^{-\mb{k}^2/4}
\exp\Bigg\{\im \sum_{x, y\in \Lambda}\frac{\beta }{2 n}U(x-y)k_x
 (\mathsf{n}_{y}\otimes \one -\one \otimes \mathsf{n}_{y})\Bigg\}.
\end{align} 
Now we obtain the assertion in the lemma. $\Box$

\begin{Prop}\label{DoubleTrace}
For each $\Phi\in \mathcal{Q}_{\beta}$,  let $\mathbb{T}_{\pm e}(\Phi_s)$
 be given by 
\begin{align}
\mathbb{T}_{\pm  e}(\Phi_s)&=\sum_{x, y\in \Lambda, |x-y|=1}t\exp\Bigg\{
\pm \im e
\int_{C_{xy}}dr\cdot \mathscr{A}(r)(\Phi_s)
\Bigg\} c_x^* c_y.
\end{align} 
One has the following:
\begin{align}
&\mathcal{Z}_{\beta, n, \vepsilon}(\mb{h})\Big/Z_{\beta+\vepsilon}\no
=&(4\pi)^{-n|\Lambda|/2}
\int_{\BbbR^{n |\Lambda|}}\prod_{j=1}^n  d\mb{k}_j
\int_{\mathcal{Q}_{ \beta+\vepsilon}}d\nu_{\beta+\vepsilon}(\Phi) 
\ex^{-\im \sum_{j=1}^n  \mb{k}_j\cdot \mb{h}}\ex^{-\sum_{j=1}^n  \mb{k}^2_j/4}
\no
&\times
\Bigg|
\Tr_{\Fock_{\mathrm{a}}}\Bigg[
\prod_{j=1}^{{n}\atop{\longrightarrow}}
\Bigg(
\prod_{(j-1)\beta/n }^{{j\beta/n}\atop{\longrightarrow} }
\ex^{\mathbb{T}_{ -e}(\Phi_s)ds}
\ex^{\im \sum_{j=1}^n \sum_{x, y\in \Lambda}\frac{\beta }{2n}k_{jx}U(x-y) \mathsf{ n}_y}
\Bigg)
\Bigg]
\Bigg|^2. \label{TraceF2}
\end{align} 
\end{Prop} 
{\it Proof.}
By the fact  $\Tr[A\otimes B]=\Tr[A] \Tr[B]$ and Lemma \ref{TraceLemma}, we obtain
\begin{align}
&\mathcal{Z}_{\beta, n, \vepsilon}(\mb{h})\Big/ Z_{\beta+\vepsilon}\no
&=(4\pi)^{-n |\Lambda|/2}
\int_{\BbbR^{n |\Lambda|}}\prod_{j=1}^n d\mb{k}_j
\int_{\mathcal{Q}_{\beta+\vepsilon}}d\nu_{\beta+\vepsilon}(\Phi) 
\ex^{-\im \sum_{j=1}^n \mb{k}_j\cdot \mb{h}}\ex^{-\sum_{j=1}^n  \mb{k}^2_j/4}
\no
&\times
\Tr_{\Fock_{\mathrm{a}}}\Bigg[
\prod_{j=1}^{{n}\atop{\longrightarrow}}
\Bigg(
\prod_{(j-1)\beta/n}^{{j\beta/n}\atop{\longrightarrow}}
\ex^{\mathbb{T}_{ -e}(\Phi_s)ds}
\ex^{+\im \sum_{j=1}^n \sum_{x, y\in \Lambda} \frac{\beta }{2n}k_{jx}U(x-y) \mathsf{n}_y}
\Bigg)
\Bigg]\no
&\times 
\Tr_{\Fock_{\mathrm{a}}}\Bigg[
\prod_{j=1}^{{n}\atop{\longrightarrow}}
\Bigg(
\prod_{(j-1)\beta/n}^{{j\beta/n}\atop{\longrightarrow}}
\ex^{\mathbb{T}_{ +e}(\Phi_s)ds}
\ex^{-\im \sum_{j=1}^n \sum_{x, y\in \Lambda}\frac{\beta }{2n}k_{jx}U(x-y) \mathsf{n}_y}
\Bigg)
\Bigg].
\end{align} 

Let $\Theta$ be a conjugation in $\Fock_{\mathrm{a}}$ defined by $\Theta c_{x_1}^* \cdots
c_{x_N}^*\Omega_{\mathrm{a}}=c_{x_1}^*\cdots c_{x_N}^*
\Omega_{\mathrm{a}}$, where $\Omega_{\mathrm{a}}$ is the Fock vaccum in $\AFock$.
Noting that $\Theta c_x \Theta =c_x$, we have 
$
\Theta \mathbb{T}_{ +e}(\Phi_s)\Theta=\mathbb{T}_{-e}(\Phi_s) $ and $\Theta \mathsf{n}_x\Theta =\mathsf{n}_x.
$
Thus it holds that 
\begin{align}
\Theta \prod_{(j-1)\beta/n}^{{j\beta/n}\atop{\longrightarrow}} \ex^{\mathbb{T}_{+e}(\Phi_s)ds}\Theta
&=\prod_{(j-1)\beta/n}^{{j\beta/n}\atop{\longrightarrow}} \ex^{\mathbb{T}_{ -e}(\Phi_s)ds},\\
\Theta \ex^{-\im \sum_{j=1}^n \sum_{x, y\in \Lambda} \frac{\beta }{2n}k_{jx} U(x-y)\mathsf{n}_y }
\Theta
&=\ex^{+\im \sum_{j=1}^n \sum_{x, y\in \Lambda} \frac{\beta }{2n}k_{jx} U(x-y)\mathsf{n}_y }.
\end{align} 
Hence using the fact $\Tr[A]=(\Tr[\Theta A \Theta])^*$, one observes
that 
\begin{align}
&\Tr_{\Fock_{\mathrm{a}}}\Bigg[
\prod_{j=1}^{{n}\atop{\longrightarrow}}
\Bigg(
\prod_{(j-1)\beta/n}^{{j\beta/n}\atop{\longrightarrow}} \ex^{\mathbb{T}_{ +e}(\Phi_s)}
\ex^{-\im \sum_{j=1}^n\sum_{x, y\in \Lambda} \frac{\beta }{2n}k_{jx}U(x-y)\mathsf{n}_y}
\Bigg)
\Bigg]\no
=&
\Bigg\{
\Tr_{\Fock_{\mathrm{a}}}\Bigg[\Theta
\prod_{j=1}^{{n}\atop{\longrightarrow}}
\Bigg(
\prod_{(j-1)\beta/n}^{{j\beta/n}\atop{\longrightarrow}} \ex^{\mathbb{T}_{ +e}(\Phi_s)}
\ex^{-\im \sum_{j=1}^n\sum_{x, y\in \Lambda} \frac{\beta }{2n}k_{jx}U(x-y)\mathsf{n}_y}
\Bigg)
\Theta
\Bigg]
\Bigg\}^*\no
=&
\Bigg\{
\Tr_{\Fock_{\mathrm{a}}}\Bigg[
\prod_{j=1}^{{n}\atop{\longrightarrow}}
\Bigg(
\prod_{(j-1)\beta/n}^{{j\beta/n}\atop{\longrightarrow}} \ex^{\mathbb{T}_{ -e}(\Phi_s)}
\ex^{+\im \sum_{j=1}^n\sum_{x, y\in \Lambda} \frac{\beta  }{2n}k_{jx}U(x-y)\mathsf{n}_y}
\Bigg)
\Bigg]
\Bigg\}^*.
\end{align} 
This completes the proof. $\Box$

\begin{Thm}\label{FiniteT}
Let $\mathcal{Z}_{\beta}(\mathbf{h})=\Tr\big[\ex^{-\beta \widehat{H}(\mathbf{h})}\big]$. Then 
for all $\mb{h}\in \BbbR^{|\Lambda|}$, 
$
\mathcal{Z}_{\beta}(\mb{h})\le \mathcal{Z}_{\beta}(\mb{0})
$ holds.
\end{Thm} 
{\it Proof.} 
By Proposition \ref{DoubleTrace},  it holds that 
$|\mathcal{Z}_{\beta, n, \vepsilon}(\mb{h})| \le
\mathcal{Z}_{\beta, n, \vepsilon}(\mb{0})$. As  $n\to \infty$,
$\mathcal{Z}_{\beta, n, \vepsilon}(\mathbf{h})$ converges to
$\mathcal{Z}_{\beta, \vepsilon}(\mathbf{h})=\Tr\big[\ex^{-\beta \widehat{H}(\mathbf{h})} \ex^{-\vepsilon d\Gamma(\check{\omega})}\big]$  by Lemma \ref{Grumm}.
Thus we have $|\mathcal{Z}_{\beta,  \vepsilon}(\mb{h})| \le
\mathcal{Z}_{\beta, \vepsilon}(\mb{0})$.
As $\vepsilon \downarrow 0$,
$\mathcal{Z}_{\beta, \vepsilon}(\mathbf{h})$ converges to
$\mathcal{Z}_{\beta}(\mathbf{h})$ by Lemma \ref{Grumm}. 
 $\Box$

\begin{lemm}\label{Grumm}
We denote by  $L^1(\mathfrak{X})$  the ideal of all trace class operators on $\mathfrak{X}$.
Let $A_n, A \in L^{\infty}(\mathfrak{X})$ and $B_n, B
 \in L^1(\mathfrak{X})$ such that $A_n$ converges to $A$ strongly and 
$\|B_n-B\|_1\to 0$ as $n\to \infty$, where $\|\cdot\|_1$ is the trace
 norm. Then $\|A_n B_n-AB\|_1\to 0$ as $n\to \infty$. 
\end{lemm} 
{\it Proof.} See \cite[Chap. 2, Example 3]{Simon2}. $\Box$
\medskip\\

\begin{coro}\label{SusInq}
For all $\mb{h}\in \BbbC^{|\Lambda|}$, we have 
\begin{align}
\Big(
\la \delta \mb{n}, \mb{U} \mb{h} \ra^*, \la \delta \mb{n}, \mb{U}\mb{h}\ra
\Big)_{\beta, \Lambda}
\le \beta^{-1} \la \mb{h}, \mb{U}\mb{h}\ra,
\end{align} 
where $\la \delta \mb{n}, \mb{U}\mb{h\ra}:=\sum_{x, y\in
 \Lambda}U(x-y)\delta n_x h_y$ and $\la \mb{h},
 \mb{U}\mb{h}\ra:=\sum_{x, y\in \Lambda}h_x^* U(x-y)h_y$.
\end{coro} 
{\it Proof.} Let $(\!(A, B )\!)_{\beta, \Lambda}$  be the Duhamel two-point function
associated with $\widehat{H}$. Then by Theorem \ref{FiniteT}, we have 
\begin{align}
\Big(\!\!\Big(
\la \mb{q}, \mb{U} \mb{h} \ra^*, \la \mb{q}, \mb{U}\mb{h}\ra
\Big)\!\!\Big)_{\beta, \Lambda} \le \beta^{-1} \la \mb{h}, \mb{U}\mb{h}\ra,
\end{align} 
where $q_x:= \mathsf{n}_x\otimes \one -\one \otimes \mathsf{n}_x$.
Since $
(\!(
\la \mb{q}, \mb{U} \mb{h} \ra^*, \la \mb{q}, \mb{U}\mb{h}\ra
)\!)_{\beta, \Lambda}=(
\la \delta \mb{n}, \mb{U} \mb{h} \ra^*, \la \delta \mb{n}, \mb{U}\mb{h}\ra
)_{\beta, \Lambda}
$, we get  the result in the corollary. $\Box$

\subsection{Completion of proof of Theorem \ref{SusUpper}}

By Corollary \ref{SusInq}, we obtain
\begin{align}
\beta\big(\widetilde{\delta n}_{-p}, \widetilde{\delta n}_p\big)_{\beta,
 \Lambda}\le \hat{U}_{\Lambda}(p)^{-1}.
\end{align} 
Since $|\hat{U}_{\Lambda}(p)- \hat{U}(p)|\to 0$ as $\ell \to \infty$
by {\bf (A. 1)}, we obtain the desired result. $\Box$

\begin{flushleft}
{\bf Acknowledgements:}
I would like to thank Prof. Arai for helpful comments.
This work was supported by KAKENHI(20554421).
\end{flushleft}

\end{document}